\documentclass[twoside,11pt]{article}

\usepackage[final]{melba}

\usepackage{amsmath,amsfonts}

\usepackage{mathabx}

\let\oldequation\equation
\let\oldendequation\endequation

\renewenvironment{align}
  {\linenomathNonumbers\oldequation}
  {\oldendequation\endlinenomath}

\melbaheading{3}{https://www.melba-journal.org/article/21200-probabilistic-dipole-inversion-for-adaptive-quantitative-susceptibility-mapping?auth_token=5HEyUtAQRW9CVoU6XAs7}{2021}{1-19}{09/2020}{03/2021}{Jinwei Zhang, Hang Zhang, Mert Sabuncu, Pascal Spincemaille, Thanh Nguyen and Yi Wang}{Medical Imaging with Deep Learning (MIDL) 2020}{Marleen de Bruijne, Tal Arbel, Ismail Ben Ayed, Hervé Lombaert}

\ShortHeadings{Probabilistic QSM}{J. Zhang, et al.}
\firstpageno{1}

\title{Probabilistic Dipole Inversion for Adaptive Quantitative Susceptibility Mapping}

\author{\name Jinwei Zhang \email jz853@cornell.edu \\  % start right after \author{, or there will be an extra space
	\addr Department of Biomedical Engineering, Cornell University, Ithaca, NY, USA \\
	\addr Department of Radiology, Weill Medical College of Cornell University, New York, NY, USA
	\AND
	\name Hang Zhang \email hz459@cornell.edu \\
	\addr Department of Electrical and Computer Engineering, Cornell University, Ithaca, NY, USA \\
	\addr Department of Radiology, Weill Medical College of Cornell University, New York, NY, USA
	\AND
	\name Mert Sabuncu \email msabuncu@cornell.edu \\
	\addr Department of Electrical and Computer Engineering, Cornell University, Ithaca, NY, USA \\
	\addr Department of Biomedical Engineering, Cornell University, Ithaca, NY, USA \\
	\addr Department of Radiology, Weill Medical College of Cornell University, New York, NY, USA
	\AND
	\name Pascal Spincemaille \email pas2018@med.cornell.edu \\
	\addr Department of Radiology, Weill Medical College of Cornell University, New York, NY, USA
	\AND
	\name Thanh Nguyen \email tdn2001@med.cornell.edu \\
	\addr Department of Radiology, Weill Medical College of Cornell University, New York, NY, USA
	\AND
	\name Yi Wang \email yiwang@med.cornell.edu \\
	\addr Department of Biomedical Engineering, Cornell University, Ithaca, NY, USA \\
	\addr Department of Radiology, Weill Medical College of Cornell University, New York, NY, USA
}

\begin{document}

% top matter
\maketitle

% abstract
\begin{abstract}%   <- trailing '%' for backward compatibility of .sty file
A learning-based posterior distribution estimation method, Probabilistic Dipole Inversion (PDI), is proposed to solve the quantitative susceptibility mapping (QSM) inverse problem in MRI with uncertainty estimation. In PDI, a deep convolutional neural network (CNN) is used to represent the multivariate Gaussian distribution as the approximate posterior distribution of susceptibility given the input measured field. Such CNN is first trained on healthy subjects via posterior density estimation, where the training dataset contains samples from the true posterior distribution. Domain adaptations are then deployed on patient datasets with new pathologies not included in pre-training, where PDI updates the pre-trained CNN's weights in an unsupervised fashion by minimizing the \emph{Kullback–Leibler} divergence between the approximate posterior distribution represented by CNN and the true posterior distribution from the likelihood distribution of a known physical model and pre-defined prior distribution. Based on our experiments, PDI provides additional uncertainty estimation compared to the conventional MAP approach, while addressing the potential issue of the pre-trained CNN when test data deviates from training.
Our code is available at~\url{https://github.com/Jinwei1209/Bayesian_QSM}.
\end{abstract}

% keywords
\begin{keywords}
Variational Inference, Uncertainty Estimation, Convolutional Neural Network, Quantitative Susceptibility Mapping
\end{keywords}

% Introduction (or first section)
\section{Introduction}
Quantitative susceptibility mapping (QSM) \citep{de2010quantitative} is an image contrast in magnetic resonance imaging (MRI) to measure the underlying tissue apparent magnetic susceptibility, which enables quantification of specific biomarkers such as iron, calcium and gadolinium \citep{wang2015quantitative}. The forward model of QSM in three dimensional image space is:
\begin{align}
   b = d \ast \chi + n 
\label{qsm_image}
\end{align}
where $\chi$ is the tissue susceptibility, $b$ is the measured local magnetic field, $d$ is the spatial dipole convolution kernel, and $n$ is the measurement noise. Dipole convolution can also be defined in k-space (Fourier space) as follows:
\begin{align}
   b = F^HDF\chi + n 
\label{qsm_kspace}
\end{align}
where $F$ is the Fourier transform operator and $D$ is the point-wise multiplication operator with the dipole kernel in k-space. The k-space formulation is more computationally efficient because of the fast Fourier transform, so Eq. \ref{qsm_kspace} is often implemented in practice. The standard deviation (SD) of the Gaussion noise $n$ is obtained by computing the variance of the least squares fit of the magnetic field $b$ from the acquired multi-echo data \citep{liu2013nonlinear}. The problem is to recover $\chi$ from $b$ due to the ill-posedness of the dipole kernel in QSM \citep{wang2015quantitative, kee2017quantitative}.

Two representative methods have been proposed to solve the QSM inverse problem. The first one is called COSMOS (calculation of susceptibility through multiple orientation sampling) \citep{liu2009calculation}. COSMOS relies on multiple orientation scans to calculate the susceptibility map with high accuracy. As a result, it has been used as the gold standard reference when developing new QSM algorithms. However, the drawback of COSMOS is that it requires at least three orientation scans, which is infeasible for clinical use. Another method called MEDI (morphology enabled dipole inversion) \citep{liu2011morphology} was proposed to solve the QSM problem with a single orientation scan. MEDI uses a morphology-related regularization term and solves the following optimization problem:
\begin{align}
    \hat{\chi} = \arg\min_{\chi}|| W(F^HDF\chi - b) ||^2_2 + \lambda|| M\nabla \chi||_1
\label{qsm_medi}
\end{align}
where $W$ is derived from the observation noise covariance matrix, $\lambda$ is the tunable parameter of weighted total variation (TV) regularization \citep{rudin1992nonlinear} with binary weighting matrix $M$ of susceptibility's spatial gradients which only penalizes regions outside the brain tissue edges in order to suppress image-space artifacts introduced by dipole inversion \citep{liu2011morphology}. Numerical optimization algorithms for solving Eq. \ref{qsm_medi} are reviewed by \cite{kee2017quantitative}. With efficient numerical solvers, MEDI generates reasonable susceptibility maps compared to COSMOS as a reference \citep{liu2011morphology} and requires only single orientation scan. As a result, MEDI has been used for clinical applications in the last decade \citep{wang2017clinical, chen2014intracranial}. 

From the Bayesian point of view, Eq. \ref{qsm_medi} belongs to the maximum a posteriori probability (MAP) estimation problem with the likelihood distribution defined as a multivariate Gaussian:
\begin{align}
p(b|\chi) = \mathcal{N}(b|F^HDF\chi, \Sigma_{b|\chi})
\label{qsm_likelihood}
\end{align}
where $n \sim \mathcal{N}(0,\Sigma_{b|\chi})$ with $\Sigma_{b|\chi}$ diagonal, and the prior distribution defined as the Laplace of the spatial gradient:  
\begin{align}
    p(\chi) \propto {\rm e}^{-\lambda\| M\nabla \chi \|_1}.
\label{qsm_prior}
\end{align}
Based on Bayes's rule, the full posterior distribution $p(\chi|b)$ given the measured local field $b$ can also be estimated in principle, which will quantify the uncertainty in the solutions delivered and may have some clinical implications. In this paper, motivated by the posterior distribution estimation problem in QSM and advances in deep learning based density estimation techniques, we introduce a set of neural network parameterized distributions to learn an approximate posterior distribution of susceptibility $\chi$ for any given $b$ with an adaptive training strategy. We validate our method on both healthy subjects and patients and show good performance of the proposed method. This paper is extended from previously published work \citep{zhang2020bayesian} at MIDL 2020. The additions include a detailed methodology section, comparisons to PDI-VI0 as another baseline in Figures 2-4 and Table 1, an experiment on multiple sclerosis patients in Figure 3, amortized versus subject-specific variational inference in Figure 5 and 6, uncertainty estimation evaluation in Figure 7, and the discussion section.

\section{Related Work}
In recent years, posterior distribution estimation in imaging inverse problems has become a new topic in medical imaging \citep{repetti2019scalable, chappell2009variational}, in which variance of a random variable is provided from posterior distribution to measure the uncertainty of the solution. However, posterior distribution estimation requires a complicated or even intractable integral from Bayes formula. Therefore, approximate inference methods are used to reduce the computational cost and intractability of the problem. Markov chain Monte Carlo (MCMC) \citep{andrieu2003introduction} and variational inference (VI) \citep{bishop2006pattern} are two classes of approximate inference approaches to the Bayesian estimation problem. In MCMC, Markov chain based sampling methods are used to generate random samples from the true posterior distribution in order to represent an empirical distribution which resembles the true distribution. MCMC is general in that it is able to achieve the exact inference given infinite computational time. However, in imaging inverse problems, the computational cost of MCMC for Bayesian estimation is often several magnitudes higher than that of the optimization method of MAP estimation, because of the curse of dimensionality \citep{pereyra2017maximum}. In addition, convergence of the Markov chain is hard to diagnose, raising concerns on the quality of the samples.

An alternative approach is to use VI, in which an approximate distribution is proposed with tractable function form and unknown parameters, and an optimization algorithm is used (for example, expectation-maximization (EM) algorithm \citep{blei2017variational}) to learn these parameters by minimizing the divergence between the true and approximate posterior distributions. After convergence, the approximate posterior distribution is expected to represent the true posterior distribution. Compared to MCMC, VI is fairly efficient as the inference problem is reduced to the optimization problem with respect to the distribution parameters. However, VI may make the model less expressive and thus lead to suboptimal performance. Although more complicated approximate function has a better representation ability in some cases, it introduces higher computational cost. Such accuracy-computation trade-off cannot be achieved easily as the inference performance depends on the tricky design of the approximate distribution form.

Due to advances in deep learning in the past few years, using deep neural network as the approximate function has become a new trend in VI. This is especially true for generative models such as variational auto-encoder (VAE) \citep{rezende2014stochastic, kingma2013auto}, in which an encoder network is built to approximate the latent variable distribution conditioned on the observed data and a decoder network is built to represent the observed data distribution conditioned on the latent variable. In addition, because of the generalization ability of a deep neural network with millions of trainable weights, amortized formulation with regularization is applied on the training dataset to learn the network weights for faster inference on the test dataset than classic VI per data, but at the expense of lower precision \citep{cremer2018inference}. As a result, this leads to a new trade-off between inference speed and amortization accuracy.

Another topic related to posterior distribution estimation with deep learning are the deep generative models trained with maximum likelihood, such as autoregressive \citep{oord2016pixel} and flow models \citep{dinh2016density}. In these models, neural network parameterized models are built to deploy tractable maximum likelihood training and generate new samples after training. If the parameterized model family is highly expressive with enough training samples, maximum likelihood training is expected to learn parameters which fit to the true data density well and generate new data with high fidelity. Autoregressive and flow models differ from VAE in that exact likelihood is evaluated in the former while approximate evaluation is applied for the latter. Such tractable inference makes training simpler but models less expressive, except for flow models which provide a combination of tractability and high expressiveness.

In this work, we propose to solve the posterior distribution estimation problem in QSM using a neural network parameterized distribution family by combining posterior density estimation from samples with posterior distribution approximation via VI for domain adaptation. Assuming multivariate Gaussian represented by a CNN as the posterior distribution of susceptibility given the input local field, a COSMOS \citep{liu2009calculation} dataset of field susceptibility pairs were used as samples from the true posterior distribution to train such CNN with an MAP loss function. Applying the likelihood in Eq. \ref{qsm_likelihood} and prior in Eq. \ref{qsm_prior}, the pre-trained CNN was adapted using VI posterior distribution approximation on different patient datasets which only contained input measured fields. Compared to MAP estimation MEDI \citep{liu2011morphology} in Eq. \ref{qsm_medi} and other deep learning QSM methods, QSMnet \citep{yoon2018quantitative} and FINE \citep{zhang2020fidelity}, the proposed method estimated the full posterior distribution of susceptibility with uncertainty quantification, while achieving domain adaptations on various datasets.

\section{Methodology}
Based on the assumption that the pattern from field $b$ to $p(\chi|b)$ is recoverable, a general distribution $p_{data}(\chi|b)$ for any given $b$ can be approximated with a learning-based approach. To accomplish that, a set of parameterized distributions $q_{\psi}(\chi|b)$ using a neural network with parameters $\psi$ are introduced and learned on a cohort of datasets including healthy subjects and patients. In this work, we assume a multivariate Gaussian distribution with diagonal covariance matrix as the approximate posterior distribution, i.e., $q_\psi(\chi|b) = \mathcal{N}(\mu_{\chi|b},\Sigma_{\chi|b})$, and use a dual-decoder network architecture (Figure 1) extended from 3D U-Net \citep{ronneberger2015u, cciccek20163d} to represent $q_{\psi}(\chi|b)$, with dual decoders' outputs representing mean $\mu_{\chi|b}$ and variance $\Sigma_{\chi|b}$ maps.

\subsection{Posterior Density Estimation}
The modeling process consists of two steps. The first step employs the COSMOS dataset. Since COSMOS provides gold standard QSM images based on multiple orientation scans, we can treat COSMOS field-susceptibility data pairs as the samples from the true posterior data distribution. Given the large amount of samples, they define an empirical distribution as follows:
\begin{align}
    \hat{p}_{data}(\chi|b) = \frac{1}{N}\sum\limits_{i=1}^N\textbf{1}[\chi = \chi_i|b = b_i]
\end{align}
where $(b_i, \chi_i)$ is the $i$-th susceptibility-field data pair sampled from $p_{data}(\chi|b)$ with total $N$ samples, and $\textbf{1}[\cdot]$ is the indicator function. We use \emph{Kullback–Leibler} (KL) divergence as the loss function to measure the distance between the empirical distribution defined by the COSMOS samples and the parameterized approximate distribution defined by the network, i.e., $KL[\hat{p}_{data}(\chi|b)||q_{\psi}(\chi|b)]$, which is equivalent to the following loss function: 
\begin{align}
    KL[\hat{p}_{data}(\chi|b)||q_{\psi}(\chi|b)] = \frac{1}{N}\sum\limits^N_{i=1}-\log q_{\psi}(\chi_i|b_i) +  H(\hat{p}_{data})
\label{kld_cosmos}
\end{align}
where the first term computes the expectation of negative log posterior density with respect to the empirical distribution and the second term is the entropy of the empirical distribution. Since the second term does not include the learnable parameters $\psi$, only the first term is used during parameter learning. Notice that training using this loss function is equivalent to maximizing the parametrized approximate posterior distribution by fitting to the dataset. Inserting $q_\psi(\chi|b) = \mathcal{N}(\mu_{\chi|b},\Sigma_{\chi|b})$ into the first term of Eq. \ref{kld_cosmos} and removing the second term of entropy, we get the loss function of posterior density estimation with the COSMOS dataset:
\begin{align}
    \frac{1}{N}\sum\limits^N_{i=1}-\log q_{\psi}(\chi_i|b_i) = \frac{1}{N}\sum\limits^N_{i=1} \frac{1}{2}(\chi_i - \mu_{\chi|b_i})^T\Sigma_{\chi|b_i}^{-1}(\chi_i - \mu_{\chi|b_i}) + \frac{1}{2} \ln |\Sigma_{\chi|b_i}|.
\label{loss_cosmos}
\end{align}
We refer to $q_\psi(\chi|b)$ trained with the COSMOS dataset as Probabilistic Dipole Inversion (PDI).

\subsection{VI Domain Adaptation}
After training with the COSMOS dataset using Eq. \ref{loss_cosmos} and obtaining the learned parameters $\psi^o$, we can simply estimate $p(\chi|b)$ as $q_{\psi^o}(\chi|b)$ given a test local field $b$. However, for a new test dataset that deviate from the COSMOS training dataset such as containing a new pathology, inferior outputs may be produced. To address this issue, $q_{\psi^o}(\chi|b)$ can be adapted by deploying VI on a subset of the new test dataset with only local field data needed in the loss function. Specifically, the pre-trained approximation network $q_{\psi}(\chi|b)$ with initial weights $\psi^o$ can be fine-tuned by minimizing the KL divergence between $p(\chi|b)$ and $q_{\psi}(\chi|b)$:
\begin{align}
\begin{split}
    & \text{ KL}[q_{\psi}(\chi|b)||p(\chi|b)]\\
     = \ &\mathbb{E}_q[\log q_{\psi}(\chi|b) - \log p(\chi|b)]\\
     = \ &\mathbb{E}_q[\log q_{\psi}(\chi|b) - \log p(\chi,b)] + \log p(b)\\\
     = \ &\text{KL}[q_{\psi}(\chi|b)||p(\chi)] - \mathbb{E}_q[\log p(b|\chi)]
\end{split}
\label{kld_derivation}
\end{align}
where the first term in the last equation imposes the approximate posterior to be similar to the prior, which works as the regularization term for training, and the second term encourages data consistency in the likelihood with the QSM foward model. Constant term $\log p(b)$ is omitted in the last equation. Inserting the prior distribution in Eq. \ref{qsm_prior} and likelihood distribution in Eq. \ref{qsm_likelihood}, the KL divergence in Eq. \ref{kld_derivation} becomes:
\begin{align}
\begin{split}
     &\text{ KL}[q_{\psi}(\chi|b)||p(\chi|b)]\\
    = \ & - \frac{1}{2}\text{ln}|\Sigma_{\chi|b}|  + \frac{1}{2K}\sum_{k=1}^{K}\lambda \| M\nabla \chi_k\|_1 + \frac{1}{2K}\sum_{k=1}^{K}(F^HDF\chi_k - b)^T\Sigma_{b|\chi}^{-1}(F^HDF\chi_k - b) 
\end{split}
\label{kld_3terms}
\end{align}
where $ - \frac{1}{2}\text{ln}|\Sigma_{\chi|b}|$ is derived from the entropy of $q_{\psi}(\chi|b)$ in $\text{KL}[q_{\psi}(\chi|b)||p(\chi)]$, $- \mathbb{E}_q[\ln p(\chi)]$ and $- \mathbb{E}_q[\log p(b|\chi)]$ are approximated through Monte Carlo (MC) sampling with $K$ samples $\chi_k's$ from $q_{\psi}(\chi|b)$. The reparameterization strategy can be used to implement back-propagation \citep{kingma2013auto}, where samples from the standard Normal distribution were used to generate samples from the predicted susceptibility distribution by scaling and translating operations, in order to make the predicted susceptibility mean and variance map learnable through back-propagation. In VI domain adaptation, Eq. \ref{kld_3terms} is minimized across the new subjects. Once trained, the adapted $q_{\psi}(\chi|b)$ can be used to predict $\mu_{\chi|b}$ and $\Sigma_{\chi|b}$ for new test subject directly, which is the so-called amortized VI. We refer to the fine-tuned approximate distribution with Eq. \ref{kld_3terms} as PDI-VI. Amortized VI can also be deployed without any COSMOS pre-training, in which only the target dataset with single orientation local field maps are needed to learn the probabilistic dipole inversion network using Eq. \ref{kld_3terms}. We refer to amortized VI without COSMOS pre-training as PDI-VI0.

The amortized formulation of VI in Eq. \ref{kld_3terms} achieves fast inference during test time compared to the classic VI per case, but potentially at the expense of suboptimal performance \citep{cremer2018inference}. This inference suboptimality can be explained as the \emph{inference gap}, which can be decomposed as follows:
\begin{align}
    \underbrace{\text{ KL}[q_{\psi^*}(\chi|b)||p(\chi|b)]}_\text{Approximation
gap} + \underbrace{\text{ KL}[q_{\psi}(\chi|b)||p(\chi|b)] - \text{ KL}[q_{\psi^*}(\chi|b)||p(\chi|b)]}_\text{Amortization gap}
\label{kld_gap}
\end{align}
where $\psi$ and $\psi^*$ are obtained by amortized and subject-specific VIs of Eq. \ref{kld_3terms}. As a result, $\text{KL}[q_{\psi}(\chi|b)||p(\chi|b)]$ is decomposed into the two terms above: the approximation gap and the amortization gap. The approximation gap is determined by the capacity of the parameterized model family $q_{\psi}(\chi|b)$ to approximate the true posterior distribution. The amortization gap is determined by the ability of the learned variational parameters $\psi$ to generalize to a new test case. Initialized with the pre-trained PDI from Eq. \ref{loss_cosmos}, we deployed and compared both amortized and subject-specific VI for QSM posterior distribution estimation.

\subsection{Relation to VAE}
The proposed VI domain adaptation strategy in Eq. \ref{kld_derivation} resembles the unsupervised variational auto-encoder \citep{kingma2013auto}. In VAE, the auto-encoder architecture is used to learn both the approximate inference network as the encoder for the latent space variable $z$ conditioned on the input data $x$, and the generative network as the decoder of data $x$ given samples of $z$. $x$ is expected to be reconstructed from $z$. Evidence lower bound (ELBO) is used to approximate the log density of data $x$ by training the encoder and decoder simultaneously, where the optimal encoder of ELBO is the true posterior distribution of $z$ given $x$, at which point the ELBO is tight and equals the log density of data $x$.

\begin{figure}[t]
  \centering
  \label{fig1}
  {\includegraphics[width=\textwidth]{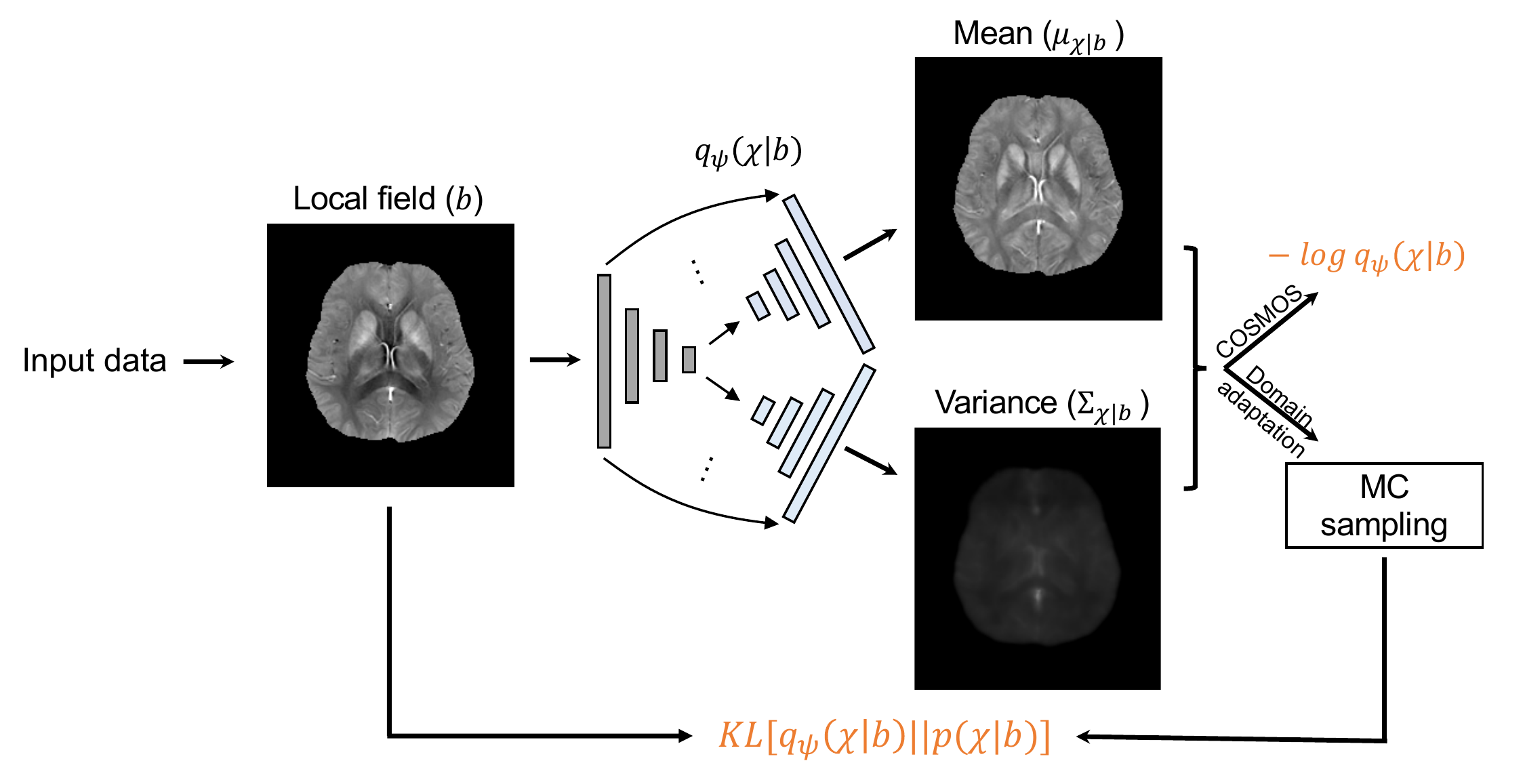}}
  {\caption{The network architecture of the proposed method. Two upsampling paths' outputs represent mean and variance maps of susceptibility. The COSMOS dataset was used to perform posterior density estimation in Eq. \ref{kld_derivation}. Domain adaptation VI with MC sampling in Eq. \ref{kld_3terms} were applied on other datasets.}}
\end{figure}

In the proposed PDI-VI strategy for QSM, the approximate posterior distribution is also a neural network "encoder" from the input field $b$ to the "latent" susceptibility $\chi$, whereas the "decoder" is no longer a neural network and does not need to be trained. Instead, this "decoder" is the likelihood distribution from the forward dipole convolution model with additive Gaussian noise in Eq. \ref{qsm_likelihood}. In addition, the prior distribution of the "latent" variable $\chi$ in Eq. \ref{qsm_prior} also comes from the domain knowledge of solving the QSM inverse problem. From physics-based likelihood and prior distributions, the same ELBO loss function in Eq. \ref{kld_derivation} is applied. Therefore, the proposed PDI-VI combines the modeling principle of distribution approximation and learning in VAE with the domain knowledge from medical physics in QSM.

\begin{figure*}[t]
  \centering
  \includegraphics[width=\textwidth]{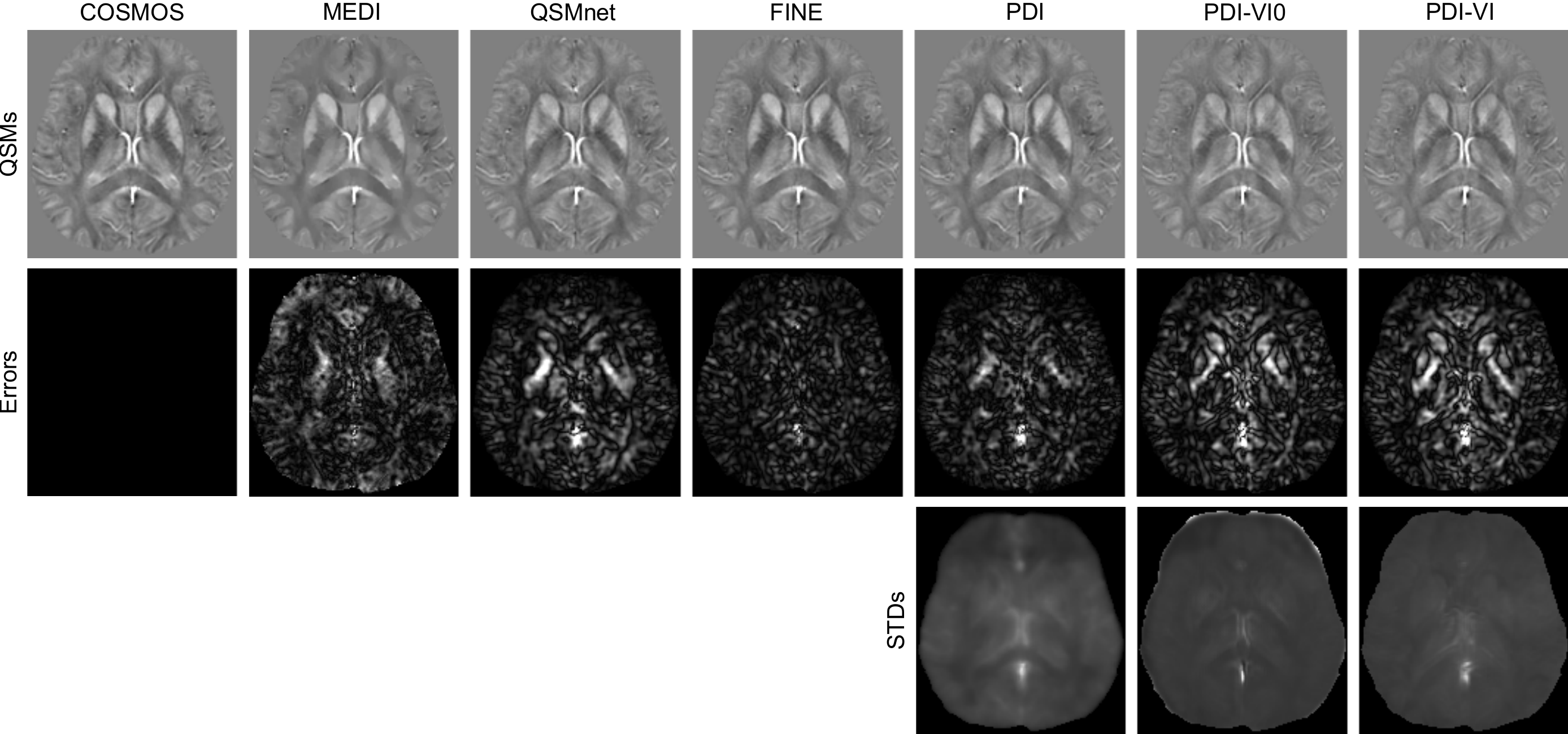}
  \caption{Reconstructions (first row, [-0.15, 0.15] ppm) and absolute error maps (second row, [0, 0.05] ppm) of one COSMOS test subject in one orientation, with COSMOS as the gold standard. FINE achieved the lowest reconstruction error, while the other methods had comparable results. SD maps of PDI, PDI-VI0 and PDI-VI (third row, [0, 0.05] ppm) showed high uncertainties at the sagittal sinus and globus pallidus, which was consistent with their error maps.}
\end{figure*}
\begin{table}
\centering
\caption{Average quantitative metrics of 10 test COSMOS brains reconstructed by different methods. FINE gave the best reconstruction at the expense of significantly increased computational time.  The other methods had comparable results.}
\begin{tabular}{cccccc} \hline
  & pSNR ($\uparrow$) & RMSE ($\downarrow$) & SSIM ($\uparrow$) & HFEN ($\downarrow$) & GPU time (s) \\ \hline
MEDI \citep{liu2012morphology} & 46.39 & 41.16 & 0.9569 & 31.30 & 17.54\\
QSMnet \citep{yoon2018quantitative} & 46.35 & 41.29 & 0.9705 & 43.31 & 0.60\\
FINE \citep{zhang2020fidelity} & 48.12 & 33.66 & 0.9789 & 31.97 & 65.42\\
PDI (Eq. \ref{loss_cosmos}) & 47.77 & 35.08 & 0.9772 & 35.17 & 0.61\\
PDI-VI0 (Eq. \ref{kld_3terms}) & 46.05 & 42.74 & 0.9704 & 42.27 & 0.61\\
PDI-VI (Eq. \ref{loss_cosmos}, then Eq. \ref{kld_3terms}) &  46.31 & 41.51 & 0.9707 & 40.58 & 0.61\\
% PDI (Eq. \ref{kld_derivation}) & 47.77 & 35.08 & 0.9772 & 35.17 & 0.61\\
\hline
\\
\end{tabular}
\end{table}

\subsection{Network Architecture}
The proposed network architecture of $q_{\psi}(\chi|b)$ is shown in Figure 1.  This network is inspired by the widely used U-Net architecture \citep{ronneberger2015u, cciccek20163d} for image-to-image mapping tasks in the biomedical deep learning field. The extension of the proposed architecture is to have one downsampling and two upsampling paths, where each upsampling path generates the mean or variance map from the same compressed feature maps. Skip concatenations between downsampling and upsampling are applied for spatial information sharing and better gradient back-propagation. Loss functions in Eqs. \ref{loss_cosmos} and \ref{kld_3terms} are used for training on COSMOS and other datasets. For the loss function in Eq. \ref{kld_3terms}, Monte Carlo sampling with reparameterization strategy is applied to stochastically optimize $q_{\psi}(\chi|b)$. The 3D convolutional kernel size is $3\times3\times3$. The numbers of filters from the highest feature level to the lowest are 32, 64, 128, 256 and 512. Batch normalization \citep{ioffe2015batch} after each convolutional layer, and max pooling operation for downsampling and deconvolutional operation for upsampling are used.

\section{Experiments}
\subsection{Data Acquisition and Preprocessing}
MRI was performed on 7 healthy subjects with 5 brain orientations using a 3T scanner (GE, Waukesha, WI) equipped with a multi-echo 3D gradient echo (GRE) sequence. The acquisition matrix was $256\times256\times48$ and voxel size was $1\times1\times3 \ \text{mm}^3$. The input local tissue field data $b$ was generated by sequentially deploying non-linear fitting across multi-echo phase data \citep{kressler2009nonlinear}, graph-cut based phase unwrapping \citep{dong2014simultaneous} and background field removal \citep{liu2011novel}. A reference QSM reconstruction was obtained using COSMOS \citep{liu2011morphology}. Two other datasets were obtained by performing single orientation GRE MRI on 9 patients with multiple sclerosis (MS) and 7 patients with intracerebral hemorrhage (ICH), which were acquired using the same scanning parameters and image processing procedures as above, except for the COSMOS reconstruction step. Data were acquired following an IRB approved protocol.

\begin{figure*}[t]
  \centering
  \includegraphics[width=\textwidth]{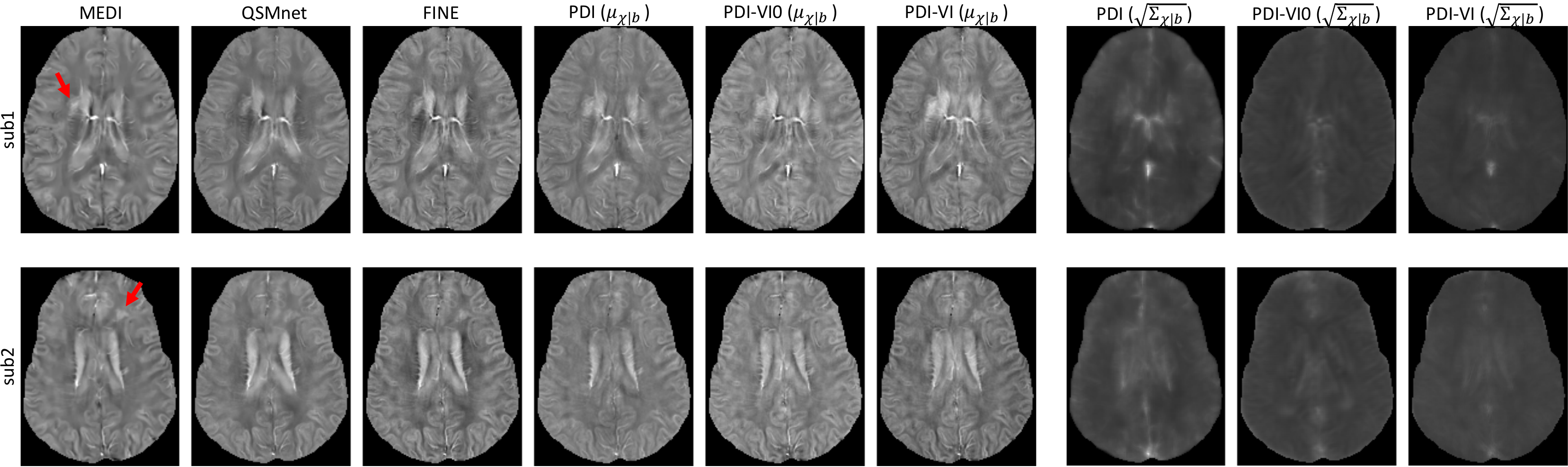}
  \caption{Two MS patient reconstructions (first six columns, [-0.15, 0.15] ppm) and SD maps (last three columns, [0, 0.05] ppm). Lesions indicated by the red arrows near the ventricle had lower susceptibility values in QSMnet and PDI, but were recovered in FINE and PDI-VI. Compared to PDI-VI, lesions reconstructed by PDI-VI0 also had lower susceptibility.}
\end{figure*}

For the COSMOS dataset, data from 4/1 subjects (20/5 brain volumes) were used as the training/validation dataset, with augmentation by in-plane rotation of $\pm15^{\circ}$. The brain volume data in the training and validation dataset was divided into 3D patches with patch size $64\times64\times32$ and extraction step $21\times21\times11$, generating $9659/2874$ patches for training/validation. Data from the remaining 2 subjects (10 brain volumes in total) were used for testing. For the MS patient dataset, data from 6/1 subjects were used as the training/validation dataset and data from the remaining 2 subjects were used for testing. For the ICH patient dataset, data from 4/1 subjects were used as the training/validation dataset and data from the remaining 2 subjects were used for testing.

\begin{figure*}[t]
  \centering
  \includegraphics[width=\textwidth]{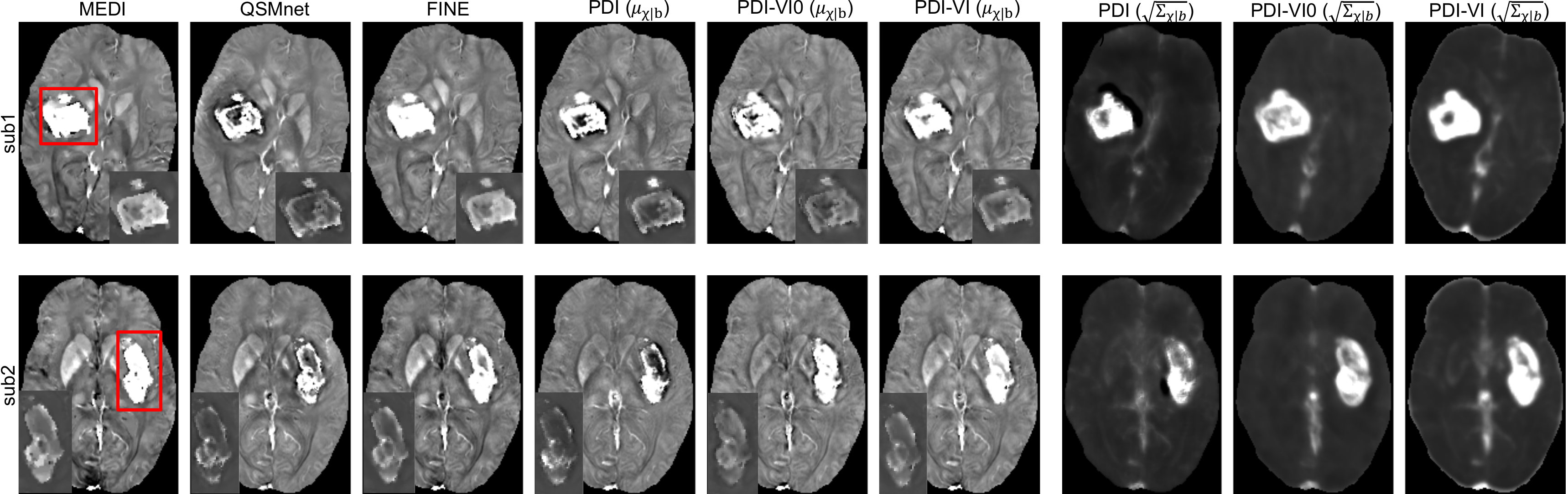}
  \caption{Two ICH patient Reconstructions (first six columns, [-0.15, 0.15] ppm) with the insets ([-0.6, 1.5] ppm) and SD maps (last three columns, [0, 0.05] ppm). Hemorrhage susceptibility was lower on QSMnet and PDI as compared to MEDI. This issue was reduced in FINE and PDI-VI. PDI-VI0 gave comparable hemorrhage reconstructions to PDI-VI. High variance inside the hemorrhage was consistent with high measured noise in the same region.}
\end{figure*}

\subsection{Implementation Details}
The loss function in Eq. \ref{loss_cosmos} was applied for posterior density estimation on the COSMOS dataset with Adam optimizer \citep{kingma2014adam} (learning rate: $10^{-3}$, Number of epochs: 60), yielding a trained network $q_{\psi^o}(\chi|b)$, denoted as PDI. Initialized with $\psi^o$, VI domain adaptations using the loss function in Eq. \ref{kld_3terms} were deployed on both MS and ICH datasets with Adam optimizer (learning rate: $10^{-3}$, Number of epochs: 100), denoted as PDI-VI. VIs using Eq. \ref{kld_3terms} and without $\psi^o$ initialization were also performed and compared for all datasets (Adam learning rate: $10^{-3}$, Number of epochs: 100), denoted as PDI-VI0. MC sampling size $K$ in VI was chosen as $5$ due to limited GPU memory. The hyperparameter $\lambda$ in Eq. \ref{kld_3terms} was chosen as 20 to balance the streaking artifact suppression and over-smoothing effect of TV regularization. While training and validation were implemented using 3D patches, whole brain volumes were fed into the network during COSMOS testing and all VI experiments. We implemented the proposed method using PyTorch (Python 3.6) on an RTX 2080Ti GPU.

\subsection{COSMOS Dataset}
For the COSMOS test dataset, we compared PDI (Eq. \ref{loss_cosmos}), PDI-VI0 (Eq. \ref{kld_3terms} without PDI pre-training) and PDI-VI (Eq. \ref{kld_3terms} with PDI pre-training) to MAP estimation MEDI \citep{liu2012morphology} and two deep learning reconstructions QSMnet \citep{yoon2018quantitative} and FINE \citep{zhang2020fidelity}. Reconstruction maps of one orientation from one test subject are shown in Figure 2. Quantitative metrics of each reconstruction method averaged among 10 test brains are shown in Table 1. FINE gave the best overall quantitative results with the expense of significantly increased computational time. The  other  methods  had  comparable  results. All deep learning methods achieved fast inference time on GPU except FINE. In Figure 2, error maps of PDI, PDI-VI0 and PDI-VI's mean outputs $\mu_{\chi|b}$ matched their SD outputs $\sqrt{\Sigma_{\chi|b}}$, with high uncertainty/error located at the sagittal sinus and globus pallidus. The SD output of PDI-VI0 and PDI-VI were sharper than PDI with lower white-grey matter variation.

\begin{figure*}[t]
  \centering
  \includegraphics[width=\textwidth]{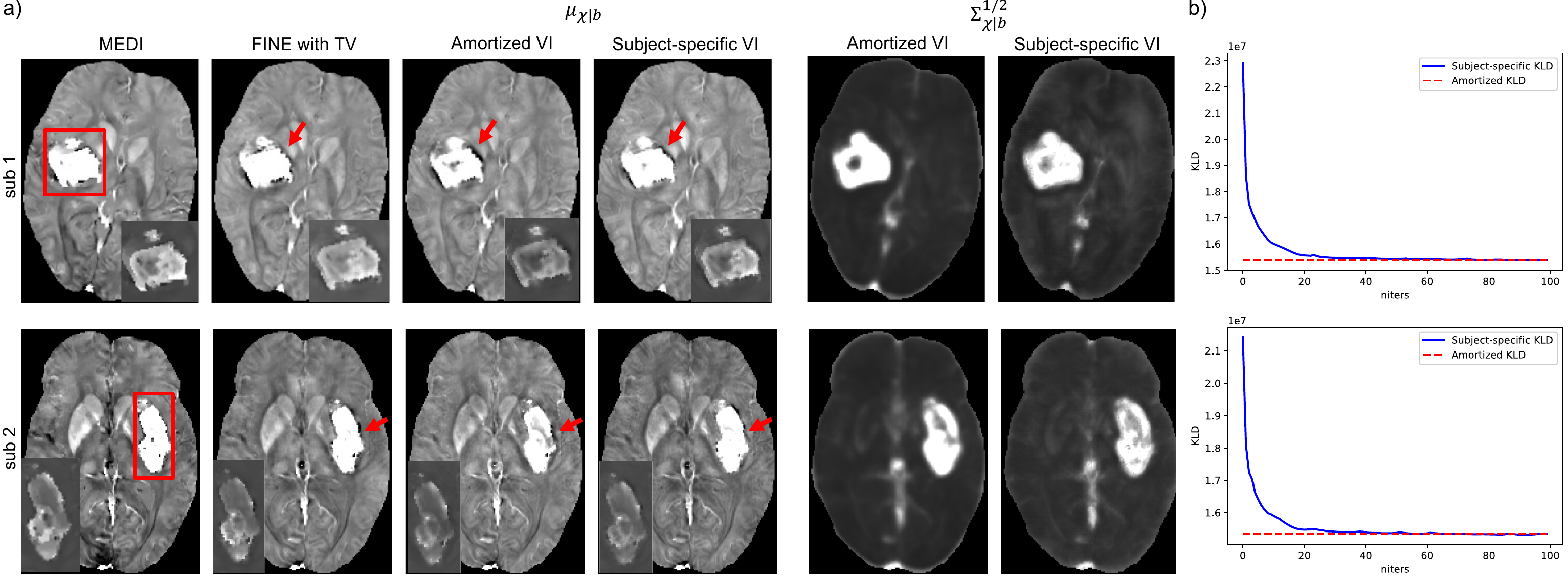}
  \caption{(a) Reconstructions ([-0.15, 0.15] ppm) with the insets ([-0.6, 1.5] ppm) and SD maps ([0, 0.05] ppm) and (b) KL divergence values of two ICH test patients using amortized and subject-specific VIs. MEDI and FINE with TV were used for comparison. Although an almost zero amortization gap (Eq. \ref{kld_gap}) was achieved by amortized VI (b) for both cases, reconstruction quality at the hemorrhage center and surrounding hemorrhage was still marginally better for subject-specific VI. FINE with TV and subject-specific VI achieve comparably image quality.}
\end{figure*}

\subsection{Patient Datasets}
The reconstruction maps of two MS patients in the test dataset are shown in Figure 3. Lesions indicated by the red arrows had susceptibility values lower in QSMnet and PDI than in MEDI, but were recovered in FINE and PDI-VI. Compared to PDI-VI, lesions reconstructed by PDI-VI0 also had lower susceptibility, which qualitatively indicated the advantage of the COSMOS dataset pre-training for PDI-VI. 

% Since these lesion patterns were not seen during COSMOS pre-training, suboptimal reconstructions could happen when applying pre-trained PDI on new test dataset deviating from the training dataset. Domain adaptation strategies of FINE and PDI-VI help correct such generalization error with different principles.

The QSMs for two ICH patients in the test dataset are shown in Figure 4. Compared to MEDI and FINE which had hyperintensity inside the hemorrhage, both QSMnet and PDI had lower susceptibility inside this region, which might result from the fact that such pathology was not encountered during training. After amortized VI domain adaptation, susceptibility value inside the hemorrhage was recovered in PDI-VI. Shadow artifacts surrounding the hemorrhage were also reduced in PDI-VI from PDI. PDI-VI0 yielded hemorrhage reconstructions that were comparable to PDI-VI. High SD map inside the reconstructed hemorrhage as shown in the last three columns of Figure 4 implied high reconstruction uncertainty of this region.

\begin{figure*}[t]
  \centering
  \includegraphics[width=\textwidth]{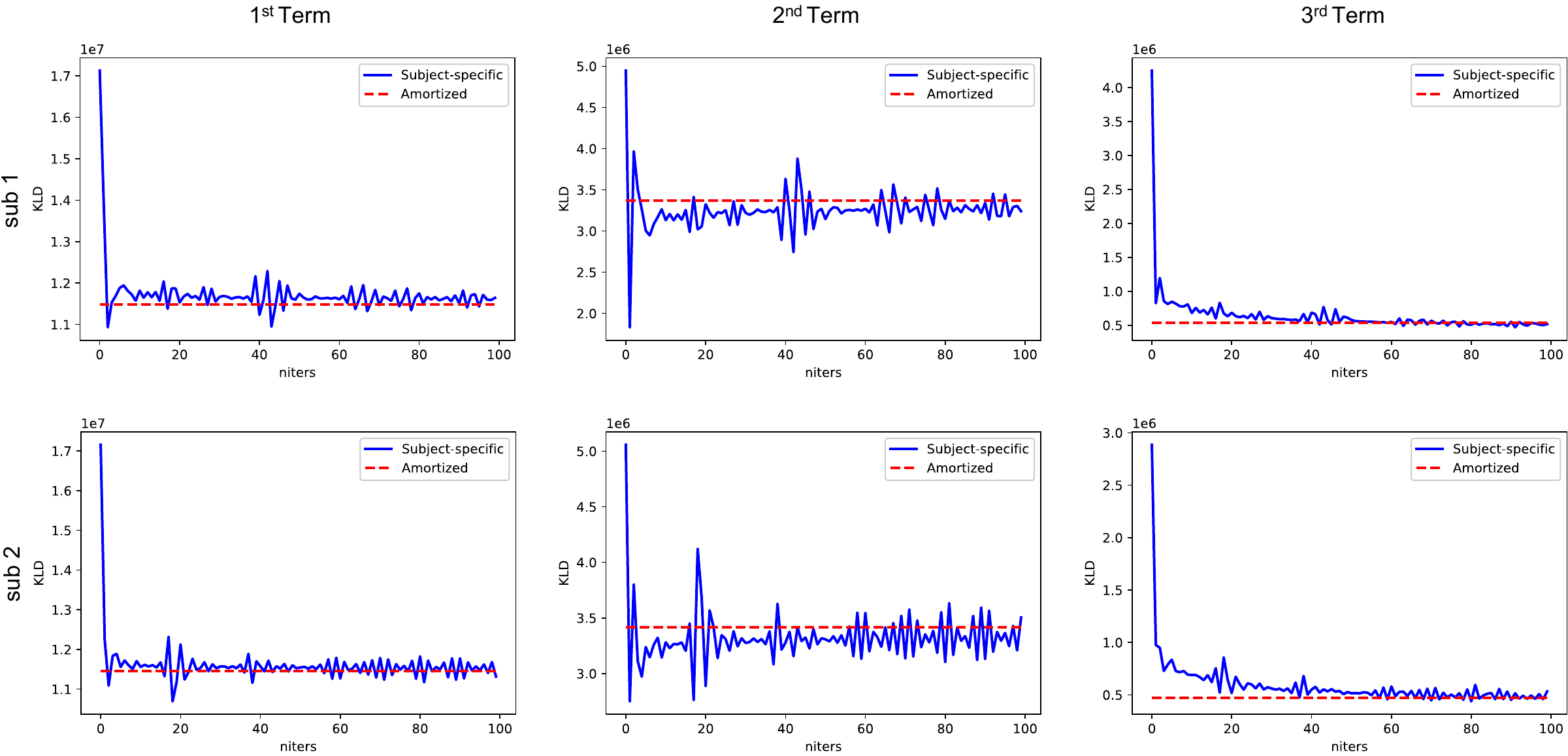}
  \caption{Value changes of three individual terms in Eq. \ref{kld_3terms} of subject-specific VI during iterations, with the value of amortized VI as a reference. The second term of TV regularization was slightly lower in subject-specific VI after convergence, while the other two terms were similar between amortized and subject-specific VIs.}
\end{figure*}

\subsection{Amortized vs Subject-specific VI}
The inference gap in Eq. \ref{kld_gap} was investigated on two ICH test cases shown in Figure 5, where subject-specific VI using Eq. \ref{kld_3terms} initialized from the weights of PDI was deployed with 100 iterations for convergence. MAP estimations in Eq. \ref{qsm_medi} of iterative reconstruction MEDI and network parametrized reconstruction FINE with TV ($\lambda = 20$, 100 iterations) were also delpoyed for comparison. As demonstrated in Figure 5a, both amortized and subject-specific VIs recovered the susceptibility value inside the hemorrhage from PDI in Figure 4. Compared to amortized VI, the susceptibility values at the center of hemorrhage (insets in Figure 5a) were further recovered and shadow artifacts surrounding the hemorrhage (red arrows in Figure 5) were reduced in subject-specific VI. In addition, subject-specific VI had similar reconstructions to MEDI and FINE with TV for both test cases, which confirmed that the mean susceptibility map by subject-specific VI equals the MAP susceptibility maps by MEDI and FINE with TV. Figure 5b shows that KL divergence of Eq. \ref{kld_3terms} during subject-specific VIs converged to the value of amortized VIs with almost zero amortization gap (Eq. \ref{kld_gap}). Figure 6 shows the value changes of three individual terms in Eq. \ref{kld_3terms} during subject-specific VI iterations, where the second term ($\frac{1}{2K}\sum_{k=1}^{K}\lambda \| M\nabla \chi_k\|_1$) was slightly lower on average than the one of amortized VI for both test cases, which might contribute to the improvement of shadow artifact reduction.

\begin{figure*}[t]
  \centering
  \includegraphics[width=\textwidth]{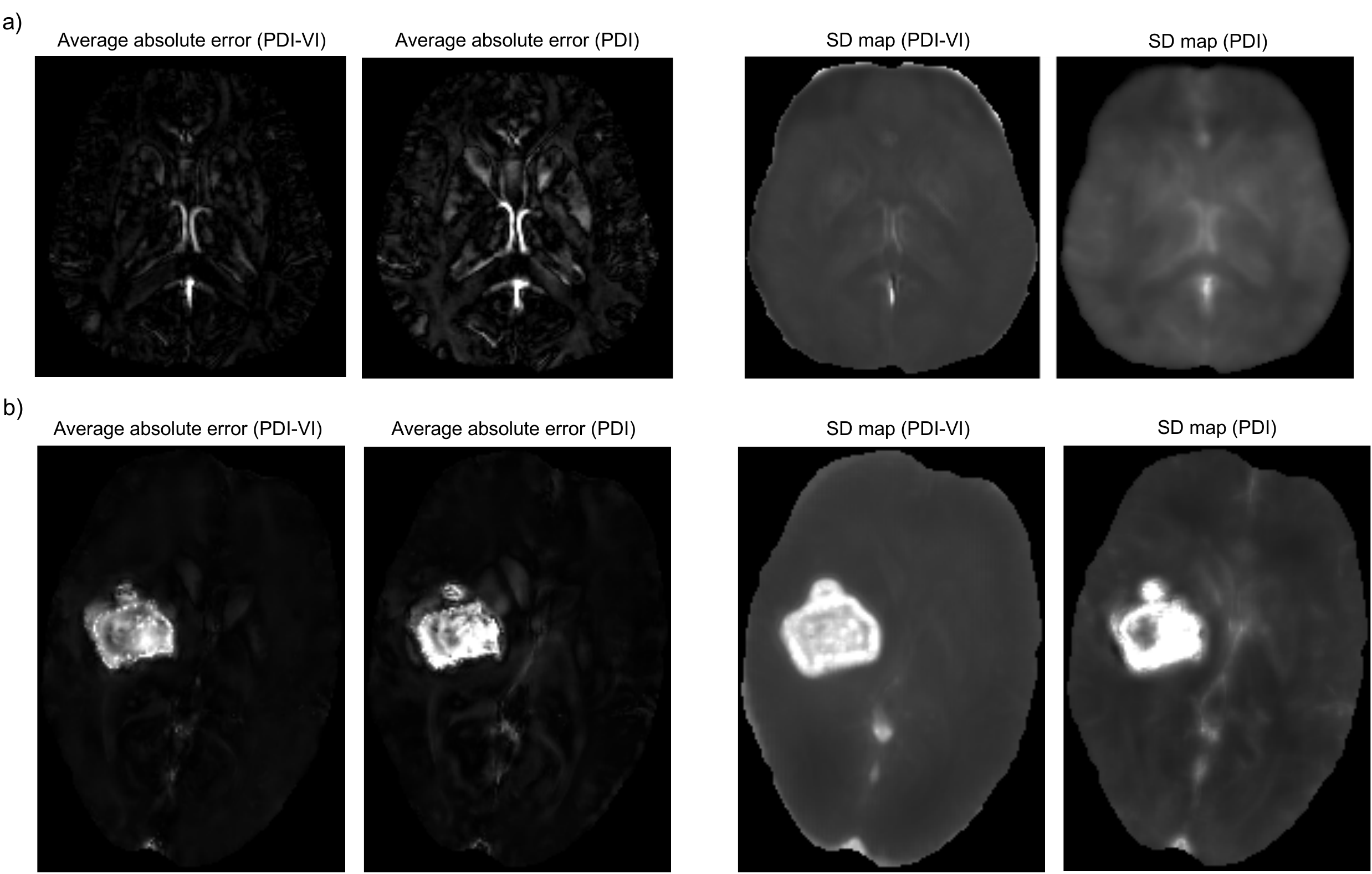}
  \caption{PDI and PDI-VI's average absolute error maps (first two columns, [0, 0.05] ppm) through simulations and predicted SD maps (last two columns, [0, 0.05] ppm) of (a) healthy and (b) hemorrhagic brains. The SD maps resembled the error maps in both cases for PDI and PDI-VI.}
\end{figure*}

\subsection{Uncertainty Map Evaluation}
To evaluate uncertainty estimation performance of the predicted SD map, absolute error maps of PDI and PDI-VI's mean predictions to the ground truth susceptibilities were computed via simulation, then correlation between susceptibility SD and error maps was examined. Local field inputs were simulated from (a) COSMOS test data in Figure 2 and (b) FINE reconstruction of the ICH patient in Figure 4a through multi-echo data synthesization with additive noise, nonlinear field fitting and phase unwrapping. Details of the simulation steps are shown in Appendix A. Such simulation was repeated 100 times to generate 100 local fields as inputs to PDI and PDI-VI. 100 mean maps of PDI and PDI-VI were predicted accordingly to compute the average absolute errors. Figure 7 shows the average absolute error maps and predicted SD maps of PDI and PDI-VI. In Figure 7a, large errors in the cerebral veins and sagittal sinus were reflected in the predicted SD maps for both PDI and PDI-VI, while in Figure 7b, large errors in the hemorrhage were also predicted in PDI and PDI-VI's SD maps, which demonstrates good correlation between the error map and the predicted SD map of the proposed method for uncertainty estimation.

\section{Discussion}
The adaptive learning strategy proposed in this paper tackles the domain adaptation challenge in medical imaging with deep learning from a probabilistic distribution refinement point of view. Since the high quality COSMOS samples are acquired only from healthy subjects, posterior density estimation with COSMOS samples may not generalize well to the patients with pathology not covered by the COSMOS dataset. As a result, even though the COSMOS pre-trained PDI performs well on COSMOS test dataset from the same distribution (Figure 2), inferior mapping happens evidenced by lower susceptibility values for lesions when applying PDI to the patients directly (Figures 3 and 4). Based on the distribution approximation principle, the pre-trained density estimation network PDI needs fine-tuning in order to fit to the patient data distribution as well. VI with KL divergence as a measure of similarity between two distributions is used for approximate distribution refinement, which helps reduce the generalization error of PDI (Figure 3 and 4). However, in terms of other domain generalizations such as different imaging resolutions, PDI-VI with KL divergence loss function for weight adjustment has not been tested and may suffer in accuracy, which will be explored in the future work. Another domain adaptation method FINE works better than PDI-VI (Figure 4) to reduce generalization error of the pre-trained network, since FINE fits to every test case by minimizing the fidelity loss, which has the major drawback of significantly increased computational time (Table 1).

The relationship between PDI-VI (Eq. \ref{kld_derivation}) and VAE \citep{kingma2013auto} is described in the methods section. The key point is that the generative network (the decoder) from latent variable to data in VAE is replaced by a physics-based likelihood model (Eq. \ref{qsm_likelihood}) in PDI-VI. This implies a general way of learning the posterior distribution of image data conditioned on the measured signal for any imaging modality, where a specific forward imaging model is used to form the "decoder" and only the "encoder" is learned with input measured signals in an unsupervised fashion like VAE. The training strategy of PDI-VI utilizes both widely available measured signals in clinic and well-defined imaging physical models to improve the reconstruction fidelity of the trained model. When gold standard reconstructions are available for training, as in the COSMOS dataset, combining direct conditional density estimation using high quality images with VI domain adaptation on measured input signals could improve the performance of VI trained on the measured signals alone (Figure 3). 

PDI defines a set of parameterized distributions using a neural network and learns these parameters from samples to approximate the true distribution, where the expressiveness of the distribution family affects their approximation ability. The network architecture (Figure 1) is inspired by 3D U-Net \citep{ronneberger2015u, cciccek20163d}, which was originally proposed for medical image segmentation tasks and has also been successfully used in deep QSM reconstructions \citep{yoon2018quantitative, zhang2020fidelity, bollmann2019deepqsm}, therefore such architecture should be expressive enough for field-to-susceptibility mapping. The COSMOS experiment indicates satisfactory image-to-image mapping ability of the proposed architecture (Figure 2 and Table 1). The simulation experiment verifies correlation between the predicted SD map and the error map, indicating reasonable uncertainty quantification of PDI and PDI-VI. Despite such merits, the choice of variational posterior form in this work is simply a Gaussian distribution with diagonal covariance matrix, which is known as the mean field approximation for modeling and calculation simplicity in classic VI. This factorized Gaussian does not consider correlation between voxels in the reconstructed susceptibility map, but in view of the forward convolution operation (Eq. \ref{qsm_image}) which aggregates the global susceptibility into the measured field at each location, taking into account the dependency between local voxels in the susceptibility map may make the variational posterior more expressive. Possible options could be improving the Gaussian posterior with a non-diagonal covariance matrix and using an autoregressive \citep{oord2016pixel} or flow-based \citep{dinh2016density} model to capture the dependency.

The prior distribution of susceptibility (Eq. \ref{qsm_prior}) used in PDI-VI comes from MEDI \citep{liu2011morphology}, where weighted TV regularization was used to suppress streaking artifacts appeared on QSM dipole inversion \citep{kee2017quantitative}. In general, the prior distribution $p(x)$ captures the density of data $x$ from a prior knowledge, where higher quality data $x$ has a higher density value. In this sense, estimating the density from sufficient data may build a more comprehensive prior distribution and therefore become more efficient to regularize the inverse problem solution. In fact, learning the prior density for MAP estimation of the imaging inverse problem has been explored by \cite{tezcan2018mr} and \cite{luo2019mri}, where VAE and PixelCNN++ \citep{salimans2017pixelcnn++} were deployed to learn the explicit prior distribution of MR images. \cite{lonning2019recurrent} proposed the recurrent inference machine by implicitly encoding the prior distribution for MAP estimation. These deep prior approaches inspire us to extend our work in the future by learning and evaluating a prior density from data and inserting them into Eqs. \ref{kld_derivation} and \ref{kld_3terms} for VI.

The inference gap (Eq. \ref{kld_gap}) summarizes two types of errors when applying the amortized inference strategy. Amortized VI has the advantage of fast inference during test time. However, it has slightly worse visual quality inside and surrounding the hemorrhage than subject-specific VI (Figure 5a). Even though an almost zero amortization gap was achieved (Figure 5b), the regularization term of KL divergence (Eq. \ref{kld_3terms}) was still better imposed in subject-specific VI, which may contribute to its better reconstruction of the hemorrhage. However, such advantage comes at a cost of extra inference time. To accelerate the inference speed of subject-specific VI, optimizing the initialization of variational parameters is useful to reduce the number of VI optimization steps. Meta-learning \citep{naik1992meta, hochreiter2001learning} may be applied to optimize the optimization process of VI per data, where a learner can be designed during pre-training to learn an inference algorithm that generalizes well to the data of interest.
 
\section{Conclusion}
In conclusion, we demonstrate a neural network parametrized distribution which yields an approximate posterior distribution of susceptibility given an input local field map for the Bayesian QSM inverse problem. The network was pre-trained on a COSMOS dataset by fitting to the empirical distribution and adapted to different domains using amortized VI. The proposed method computes adaptive reconstructions of susceptibility together with an uncertainty estimation. Future work will include building a more expressive posterior distribution family, learning a deep prior density for regularization and optimizing the subject-specific VI algorithm using meta-learning.

%%%%%%%%%%%%%%%%%%%%%%%%%%%%%%%%%%%%%%%%%%%%%%%%%%%%%%%%%%%%%%%%%%%%%%%
% Mandatory Sections. Please complete, especially for final publication
%%%%%%%%%%%%%%%%%%%%%%%%%%%%%%%%%%%%%%%%%%%%%%%%%%%%%%%%%%%%%%%%%%%%%%%

% Acknowledgements. 
% Please include any funding, intellectual contributions not included in the authorship, and any other acknowledgements. 
\acks{The authors would like to thank Adrian Dalca for useful feedback and discussions.
This research was supported in part by National Institute of Health (R01NS090464, R01NS095562, R01NS105144, R01DK116126, R01CA181566, S10OD021782, 1R21AG050122, R01LM012719 and R01AG053949), National Science Foundation (1748377 and 1707312) and National Multiple Sclerosis Society (RR-1602-07671).}

% Ethical Standards.
% Please edit with the appropriate ethics considerations for your work. Include any pertinent IRB information, etc.
% 
% Please note that the submission requirements included:
% The work presented must follow appropriate ethical standards in conducting research and writing the manuscript, following all applicable laws and regulations regarding treatment of animals or human subjects.
\ethics{The work follows appropriate ethical standards in conducting research and writing the manuscript. Data were acquired following an IRB approved protocol.}

% Conflict of Interest
% Declaration of possible conflicts of interest: Authors must disclose any financial, organisational, commercial or personal conflicts of interest that might bias their work.  
% If no conflicts, please say "We declare we don't have conflicts of interest."
\coi{We declare we don't have conflicts of interest.}

% Manual newpage inserted to improve layout of sample file - not
% needed in general before appendices/bibliography.
% \newpage

\appendix % optional
\section*{Appendix A.}
In this appendix we show the steps of simulation in section 4.6:
\begin{itemize}
  \item Synthesize local magnetic field data $b_{syn}$ from COSMOS gold standard susceptibility $\chi_{COSMOS}$ using dipole convolution model:
  $$ b_{syn} = \chi_{COSMOS} * d$$

  \item Synthesize multi-echo MR images $S_j$ from $b_{syn}$ (above), $R2^*$ ($T2^*$ decay rate) and $M_0$ (water) using forward physical model: 
  $$S_j = M_0e^{-R2^* \cdot t_j}e^{i(\phi_0 + b_{syn} \cdot t_j)} + n_j, $$ 
  where $t_j$ is the echo time of the $j$-th echo, $n_j$ is the i.i.d. Gaussian noise on real and imag parts for all voxels and initial phase $\phi_0 = 0$ is assumed for all voxels.
  
 \item Deploy nonlinear field fitting and spatial phase unwrapping to estimate the noisy local field data.

\end{itemize}

\vskip 0.2in
\bibliography{sample}

\begin{thebibliography}{38}
\providecommand{\natexlab}[1]{#1}
\providecommand{\url}[1]{\texttt{#1}}
\expandafter\ifx\csname urlstyle\endcsname\relax
  \providecommand{\doi}[1]{doi: #1}\else
  \providecommand{\doi}{doi: \begingroup \urlstyle{rm}\Url}\fi

\bibitem[Andrieu et~al.(2003)Andrieu, De~Freitas, Doucet, and
  Jordan]{andrieu2003introduction}
Christophe Andrieu, Nando De~Freitas, Arnaud Doucet, and Michael~I Jordan.
\newblock An introduction to mcmc for machine learning.
\newblock \emph{Machine learning}, 50\penalty0 (1-2):\penalty0 5--43, 2003.

\bibitem[Bishop(2006)]{bishop2006pattern}
Christopher~M Bishop.
\newblock \emph{Pattern recognition and machine learning}.
\newblock springer, 2006.

\bibitem[Blei et~al.(2017)Blei, Kucukelbir, and McAuliffe]{blei2017variational}
David~M Blei, Alp Kucukelbir, and Jon~D McAuliffe.
\newblock Variational inference: A review for statisticians.
\newblock \emph{Journal of the American Statistical Association}, 112\penalty0
  (518):\penalty0 859--877, 2017.

\bibitem[Bollmann et~al.(2019)Bollmann, Rasmussen, Kristensen, Blendal,
  {\O}stergaard, Plocharski, O'Brien, Langkammer, Janke, and
  Barth]{bollmann2019deepqsm}
Steffen Bollmann, Kasper Gade~B{\o}tker Rasmussen, Mads Kristensen,
  Rasmus~Guldhammer Blendal, Lasse~Riis {\O}stergaard, Maciej Plocharski,
  Kieran O'Brien, Christian Langkammer, Andrew Janke, and Markus Barth.
\newblock Deepqsm-using deep learning to solve the dipole inversion for
  quantitative susceptibility mapping.
\newblock \emph{Neuroimage}, 195:\penalty0 373--383, 2019.

\bibitem[Chappell et~al.(2009)Chappell, Groves, Whitcher, and
  Woolrich]{chappell2009variational}
Michael~A Chappell, Adrian~R Groves, Brandon Whitcher, and Mark~W Woolrich.
\newblock Variational bayesian inference for a nonlinear forward model.
\newblock \emph{IEEE Transactions on Signal Processing}, 57\penalty0
  (1):\penalty0 223--236, 2009.

\bibitem[Chen et~al.(2014)Chen, Zhu, Kovanlikaya, Kovanlikaya, Liu, Wang,
  Salustri, and Wang]{chen2014intracranial}
Weiwei Chen, Wenzhen Zhu, IIhami Kovanlikaya, Arzu Kovanlikaya, Tian Liu, Shuai
  Wang, Carlo Salustri, and Yi~Wang.
\newblock Intracranial calcifications and hemorrhages: characterization with
  quantitative susceptibility mapping.
\newblock \emph{Radiology}, 270\penalty0 (2):\penalty0 496--505, 2014.

\bibitem[{\c{C}}i{\c{c}}ek et~al.(2016){\c{C}}i{\c{c}}ek, Abdulkadir, Lienkamp,
  Brox, and Ronneberger]{cciccek20163d}
{\"O}zg{\"u}n {\c{C}}i{\c{c}}ek, Ahmed Abdulkadir, Soeren~S Lienkamp, Thomas
  Brox, and Olaf Ronneberger.
\newblock 3d u-net: learning dense volumetric segmentation from sparse
  annotation.
\newblock In \emph{International conference on medical image computing and
  computer-assisted intervention}, pages 424--432. Springer, 2016.

\bibitem[Cremer et~al.(2018)Cremer, Li, and Duvenaud]{cremer2018inference}
Chris Cremer, Xuechen Li, and David Duvenaud.
\newblock Inference suboptimality in variational autoencoders.
\newblock \emph{arXiv preprint arXiv:1801.03558}, 2018.

\bibitem[de~Rochefort et~al.(2010)de~Rochefort, Liu, Kressler, Liu,
  Spincemaille, Lebon, Wu, and Wang]{de2010quantitative}
Ludovic de~Rochefort, Tian Liu, Bryan Kressler, Jing Liu, Pascal Spincemaille,
  Vincent Lebon, Jianlin Wu, and Yi~Wang.
\newblock Quantitative susceptibility map reconstruction from mr phase data
  using bayesian regularization: validation and application to brain imaging.
\newblock \emph{Magnetic Resonance in Medicine: An Official Journal of the
  International Society for Magnetic Resonance in Medicine}, 63\penalty0
  (1):\penalty0 194--206, 2010.

\bibitem[Dinh et~al.(2016)Dinh, Sohl-Dickstein, and Bengio]{dinh2016density}
Laurent Dinh, Jascha Sohl-Dickstein, and Samy Bengio.
\newblock Density estimation using real nvp.
\newblock \emph{arXiv preprint arXiv:1605.08803}, 2016.

\bibitem[Dong et~al.(2014)Dong, Liu, Chen, Zhou, Dimov, Raj, Cheng,
  Spincemaille, and Wang]{dong2014simultaneous}
Jianwu Dong, Tian Liu, Feng Chen, Dong Zhou, Alexey Dimov, Ashish Raj, Qiang
  Cheng, Pascal Spincemaille, and Yi~Wang.
\newblock Simultaneous phase unwrapping and removal of chemical shift (spurs)
  using graph cuts: application in quantitative susceptibility mapping.
\newblock \emph{IEEE transactions on medical imaging}, 34\penalty0
  (2):\penalty0 531--540, 2014.

\bibitem[Hochreiter et~al.(2001)Hochreiter, Younger, and
  Conwell]{hochreiter2001learning}
Sepp Hochreiter, A~Steven Younger, and Peter~R Conwell.
\newblock Learning to learn using gradient descent.
\newblock In \emph{International Conference on Artificial Neural Networks},
  pages 87--94. Springer, 2001.

\bibitem[Ioffe and Szegedy(2015)]{ioffe2015batch}
Sergey Ioffe and Christian Szegedy.
\newblock Batch normalization: Accelerating deep network training by reducing
  internal covariate shift.
\newblock \emph{arXiv preprint arXiv:1502.03167}, 2015.

\bibitem[Kee et~al.(2017)Kee, Liu, Zhou, Dimov, Cho, De~Rochefort, Seo, and
  Wang]{kee2017quantitative}
Youngwook Kee, Zhe Liu, Liangdong Zhou, Alexey Dimov, Junghun Cho, Ludovic
  De~Rochefort, Jin~Keun Seo, and Yi~Wang.
\newblock Quantitative susceptibility mapping (qsm) algorithms: mathematical
  rationale and computational implementations.
\newblock \emph{IEEE Transactions on Biomedical Engineering}, 64\penalty0
  (11):\penalty0 2531--2545, 2017.

\bibitem[Kingma and Ba(2014)]{kingma2014adam}
Diederik~P Kingma and Jimmy Ba.
\newblock Adam: A method for stochastic optimization.
\newblock \emph{arXiv preprint arXiv:1412.6980}, 2014.

\bibitem[Kingma and Welling(2013)]{kingma2013auto}
Diederik~P Kingma and Max Welling.
\newblock Auto-encoding variational bayes.
\newblock \emph{arXiv preprint arXiv:1312.6114}, 2013.

\bibitem[Kressler et~al.(2009)Kressler, De~Rochefort, Liu, Spincemaille, Jiang,
  and Wang]{kressler2009nonlinear}
Bryan Kressler, Ludovic De~Rochefort, Tian Liu, Pascal Spincemaille, Quan
  Jiang, and Yi~Wang.
\newblock Nonlinear regularization for per voxel estimation of magnetic
  susceptibility distributions from mri field maps.
\newblock \emph{IEEE transactions on medical imaging}, 29\penalty0
  (2):\penalty0 273--281, 2009.

\bibitem[Liu et~al.(2012)Liu, Liu, de~Rochefort, Ledoux, Khalidov, Chen,
  Tsiouris, Wisnieff, Spincemaille, Prince, et~al.]{liu2012morphology}
Jing Liu, Tian Liu, Ludovic de~Rochefort, James Ledoux, Ildar Khalidov, Weiwei
  Chen, A~John Tsiouris, Cynthia Wisnieff, Pascal Spincemaille, Martin~R
  Prince, et~al.
\newblock Morphology enabled dipole inversion for quantitative susceptibility
  mapping using structural consistency between the magnitude image and the
  susceptibility map.
\newblock \emph{Neuroimage}, 59\penalty0 (3):\penalty0 2560--2568, 2012.

\bibitem[Liu et~al.(2009)Liu, Spincemaille, De~Rochefort, Kressler, and
  Wang]{liu2009calculation}
Tian Liu, Pascal Spincemaille, Ludovic De~Rochefort, Bryan Kressler, and
  Yi~Wang.
\newblock Calculation of susceptibility through multiple orientation sampling
  (cosmos): a method for conditioning the inverse problem from measured
  magnetic field map to susceptibility source image in mri.
\newblock \emph{Magnetic Resonance in Medicine: An Official Journal of the
  International Society for Magnetic Resonance in Medicine}, 61\penalty0
  (1):\penalty0 196--204, 2009.

\bibitem[Liu et~al.(2011{\natexlab{a}})Liu, Khalidov, de~Rochefort,
  Spincemaille, Liu, Tsiouris, and Wang]{liu2011novel}
Tian Liu, Ildar Khalidov, Ludovic de~Rochefort, Pascal Spincemaille, Jing Liu,
  A~John Tsiouris, and Yi~Wang.
\newblock A novel background field removal method for mri using projection onto
  dipole fields.
\newblock \emph{NMR in Biomedicine}, 24\penalty0 (9):\penalty0 1129--1136,
  2011{\natexlab{a}}.

\bibitem[Liu et~al.(2011{\natexlab{b}})Liu, Liu, De~Rochefort, Spincemaille,
  Khalidov, Ledoux, and Wang]{liu2011morphology}
Tian Liu, Jing Liu, Ludovic De~Rochefort, Pascal Spincemaille, Ildar Khalidov,
  James~Robert Ledoux, and Yi~Wang.
\newblock Morphology enabled dipole inversion (medi) from a single-angle
  acquisition: comparison with cosmos in human brain imaging.
\newblock \emph{Magnetic resonance in medicine}, 66\penalty0 (3):\penalty0
  777--783, 2011{\natexlab{b}}.

\bibitem[Liu et~al.(2013)Liu, Wisnieff, Lou, Chen, Spincemaille, and
  Wang]{liu2013nonlinear}
Tian Liu, Cynthia Wisnieff, Min Lou, Weiwei Chen, Pascal Spincemaille, and
  Yi~Wang.
\newblock Nonlinear formulation of the magnetic field to source relationship
  for robust quantitative susceptibility mapping.
\newblock \emph{Magnetic resonance in medicine}, 69\penalty0 (2):\penalty0
  467--476, 2013.

\bibitem[L{\o}nning et~al.(2019)L{\o}nning, Putzky, Sonke, Reneman, Caan, and
  Welling]{lonning2019recurrent}
Kai L{\o}nning, Patrick Putzky, Jan-Jakob Sonke, Liesbeth Reneman, Matthan~WA
  Caan, and Max Welling.
\newblock Recurrent inference machines for reconstructing heterogeneous mri
  data.
\newblock \emph{Medical image analysis}, 53:\penalty0 64--78, 2019.

\bibitem[Luo et~al.(2019)Luo, Zhao, Jiang, and Cao]{luo2019mri}
GuanXiong Luo, Na~Zhao, Wenhao Jiang, and Peng Cao.
\newblock Mri reconstruction using deep bayesian inference.
\newblock \emph{arXiv preprint arXiv:1909.01127}, 2019.

\bibitem[Naik and Mammone(1992)]{naik1992meta}
Devang~K Naik and Richard~J Mammone.
\newblock Meta-neural networks that learn by learning.
\newblock In \emph{[Proceedings 1992] IJCNN International Joint Conference on
  Neural Networks}, volume~1, pages 437--442. IEEE, 1992.

\bibitem[Oord et~al.(2016)Oord, Kalchbrenner, and Kavukcuoglu]{oord2016pixel}
Aaron van~den Oord, Nal Kalchbrenner, and Koray Kavukcuoglu.
\newblock Pixel recurrent neural networks.
\newblock \emph{arXiv preprint arXiv:1601.06759}, 2016.

\bibitem[Pereyra(2017)]{pereyra2017maximum}
Marcelo Pereyra.
\newblock Maximum-a-posteriori estimation with bayesian confidence regions.
\newblock \emph{SIAM Journal on Imaging Sciences}, 10\penalty0 (1):\penalty0
  285--302, 2017.

\bibitem[Repetti et~al.(2019)Repetti, Pereyra, and Wiaux]{repetti2019scalable}
Audrey Repetti, Marcelo Pereyra, and Yves Wiaux.
\newblock Scalable bayesian uncertainty quantification in imaging inverse
  problems via convex optimization.
\newblock \emph{SIAM Journal on Imaging Sciences}, 12\penalty0 (1):\penalty0
  87--118, 2019.

\bibitem[Rezende et~al.(2014)Rezende, Mohamed, and
  Wierstra]{rezende2014stochastic}
Danilo~Jimenez Rezende, Shakir Mohamed, and Daan Wierstra.
\newblock Stochastic backpropagation and approximate inference in deep
  generative models.
\newblock \emph{arXiv preprint arXiv:1401.4082}, 2014.

\bibitem[Ronneberger et~al.(2015)Ronneberger, Fischer, and
  Brox]{ronneberger2015u}
Olaf Ronneberger, Philipp Fischer, and Thomas Brox.
\newblock U-net: Convolutional networks for biomedical image segmentation.
\newblock In \emph{International Conference on Medical image computing and
  computer-assisted intervention}, pages 234--241. Springer, 2015.

\bibitem[Rudin et~al.(1992)Rudin, Osher, and Fatemi]{rudin1992nonlinear}
Leonid~I Rudin, Stanley Osher, and Emad Fatemi.
\newblock Nonlinear total variation based noise removal algorithms.
\newblock \emph{Physica D: nonlinear phenomena}, 60\penalty0 (1-4):\penalty0
  259--268, 1992.

\bibitem[Salimans et~al.(2017)Salimans, Karpathy, Chen, and
  Kingma]{salimans2017pixelcnn++}
Tim Salimans, Andrej Karpathy, Xi~Chen, and Diederik~P Kingma.
\newblock Pixelcnn++: Improving the pixelcnn with discretized logistic mixture
  likelihood and other modifications.
\newblock \emph{arXiv preprint arXiv:1701.05517}, 2017.

\bibitem[Tezcan et~al.(2018)Tezcan, Baumgartner, Luechinger, Pruessmann, and
  Konukoglu]{tezcan2018mr}
Kerem~C Tezcan, Christian~F Baumgartner, Roger Luechinger, Klaas~P Pruessmann,
  and Ender Konukoglu.
\newblock Mr image reconstruction using deep density priors.
\newblock \emph{IEEE transactions on medical imaging}, 2018.

\bibitem[Wang and Liu(2015)]{wang2015quantitative}
Yi~Wang and Tian Liu.
\newblock Quantitative susceptibility mapping (qsm): decoding mri data for a
  tissue magnetic biomarker.
\newblock \emph{Magnetic resonance in medicine}, 73\penalty0 (1):\penalty0
  82--101, 2015.

\bibitem[Wang et~al.(2017)Wang, Spincemaille, Liu, Dimov, Deh, Li, Zhang, Yao,
  Gillen, Wilman, et~al.]{wang2017clinical}
Yi~Wang, Pascal Spincemaille, Zhe Liu, Alexey Dimov, Kofi Deh, Jianqi Li, Yan
  Zhang, Yihao Yao, Kelly~M Gillen, Alan~H Wilman, et~al.
\newblock Clinical quantitative susceptibility mapping (qsm): biometal imaging
  and its emerging roles in patient care.
\newblock \emph{Journal of Magnetic Resonance Imaging}, 46\penalty0
  (4):\penalty0 951--971, 2017.

\bibitem[Yoon et~al.(2018)Yoon, Gong, Chatnuntawech, Bilgic, Lee, Jung, Ko,
  Jung, Setsompop, Zaharchuk, et~al.]{yoon2018quantitative}
Jaeyeon Yoon, Enhao Gong, Itthi Chatnuntawech, Berkin Bilgic, Jingu Lee, Woojin
  Jung, Jingyu Ko, Hosan Jung, Kawin Setsompop, Greg Zaharchuk, et~al.
\newblock Quantitative susceptibility mapping using deep neural network:
  Qsmnet.
\newblock \emph{Neuroimage}, 179:\penalty0 199--206, 2018.

\bibitem[Zhang et~al.(2020{\natexlab{a}})Zhang, Liu, Zhang, Zhang,
  Spincemaille, Nguyen, Sabuncu, and Wang]{zhang2020fidelity}
Jinwei Zhang, Zhe Liu, Shun Zhang, Hang Zhang, Pascal Spincemaille, Thanh~D
  Nguyen, Mert~R Sabuncu, and Yi~Wang.
\newblock Fidelity imposed network edit (fine) for solving ill-posed image
  reconstruction.
\newblock \emph{NeuroImage}, page 116579, 2020{\natexlab{a}}.

\bibitem[Zhang et~al.(2020{\natexlab{b}})Zhang, Zhang, Sabuncu, Spincemaille,
  Nguyen, and Wang]{zhang2020bayesian}
Jinwei Zhang, Hang Zhang, Mert Sabuncu, Pascal Spincemaille, Thanh Nguyen, and
  Yi~Wang.
\newblock Bayesian learning of probabilistic dipole inversion for quantitative
  susceptibility mapping.
\newblock In \emph{Medical Imaging with Deep Learning}, 2020{\natexlab{b}}.

\end{thebibliography}

\end{document}